\documentclass{PoS}

\newcommand{\mev}{\,{\rm MeV}}
\newcommand{\cm}{\,{\rm cm}}

\newcommand{\etal}{{\it et al.}~}
\newcommand{\del}{\partial}
\newcommand{\qbar}{\overline{q}}

\title{Strange quark content of the nucleon and dark matter searches}

\ShortTitle{Strange quark content of the nucleon and dark matter searches}

\author{\speaker{R.~D.~Young}\\
        Special Research Centre for the Subatomic Structure of Matter (CSSM)\\
        and ARC Centre of Excellence in Particle Physics at the Terascale (CoEPP),\\
        School of Chemistry and Physics, University of Adelaide, SA 5005, Australia.\\
        E-mail: \email{ross.young@adelaide.edu.au}}

      \abstract{
        The strange quark scalar content plays an important role in
        both the description of nucleon structure and in the
        determination of dark matter direct detection cross
        sections. As a measure of the strange-quark contribution to
        the nucleon mass, the strange-quark sigma term ($\sigma_s$)
        provides important insight into the nature of mass generation
        in QCD. The phenomenological determination of $\sigma_s$ exhibits
        a wide range of variation, with values suggesting that the
        strange quark contributes anywhere between $0$ and more than $30\%$ of
        the nucleon mass.  In the context of dark matter searches,
        coupled with relatively large Higgs coupling to strangeness,
        this variation dominates the uncertainty in predicted cross
        sections for a large class of dark matter models.  Here we
        report on the recent results in lattice QCD, which are now
        giving a far more precise determination of $\sigma_s$ than can
        be inferred from phenomenology. As a consequence, the lattice
        determinations of $\sigma_s$ can now dramatically reduce the
        uncertainty in dark matter cross sections associated with the
        hadronic matrix elements.}

\FullConference{The 30 International Symposium on Lattice Field Theory - Lattice 2012,\\
		June 24-29, 2012\\
		Cairns, Australia}

\begin{document}

\section{Introduction}
The modern picture of the universe suggests that the ordinary matter
component in the energy composition of the universe is only about 4\%
--- with the remainder comprised of some form of cold, ``dark
matter'', which clusters around ordinary matter and an even larger ``dark
energy'' component.  While little is known about the physical nature
of dark energy, there is strong evidence that suggests we are nearing
the discovery phase in the identification of dark matter. Supposing
dark matter to take a particle-like form, the relevant mass scale for
such particles are most likely to be within reach of the LHC.

For a general spin-independent interaction of a WIMP with a nucleus,
the low-energy limit reduces to a scalar contact interaction, and
hence sensitive to the $\qbar q$ matrix elements within a nucleon. And
for Higgs-dominated exchange, these couplings are proportional to the
corresponding quark mass and are hence sensitive to the nucleon sigma
terms. The sigma terms therefore become of principal uncertainty in
the predicting the cross sections associated with any candidate model
of dark matter. The significance of the uncertainty in these hadronic
matrix elements has been highlighted in a range of dark matter models,
see for example
Refs.~\cite{Bottino:1999ei,Ellis:2005mb,Ellis:2008hf,Bottino:2008mf,Hill:2011be}.

Before going on to discuss the sigma terms, and the strangeness scalar
content, it is worth recapping some recent advances in resolving the
strange quark vector form factors in the nucleon. Early estimates of
the strangeness electromagnetic currents had suggested that they could
be relatively large \cite{Jaffe:1989mj}. After a dedicated
experimental effort, parity-violating electron scattering measurements
\cite{Maas:2004dh,Armstrong:2005hs,Acha:2006my} have revealed that the
strange quarks contribute much less than originally suggested. This
finding is also supported by lattice-based phenomenological estimates
\cite{Leinweber:2004tc,Leinweber:2006ug} and recent direct lattice QCD
simulation results \cite{Doi:2009sq,Babich:2010at}. For a recent
review of these latest revelations, see Ref.~\cite{Paschke:2011zz}.

After setting some general notation, the phenomenological
determination of the sigma terms are reviewed in Section 2; a summary
of the latest lattice QCD results are reported in Section 3; the
significance of these results in the context of dark matter searches
are discussed in Section 4; follwed with a summary in Section 5.

\subsection{Notation}
Of primary interest here are the scalar nucleon matrix elements, where
we'll use the notation for the light- and strange-quark matrix
elements
\begin{equation}
\sigma_l\equiv m_l\langle N|\overline{u}u+\overline{d}d|N\rangle\,,\qquad
\sigma_s\equiv m_s\langle N|\overline{s}s|N\rangle\,,
\end{equation}
with the average light-quark mass given by $m_l\equiv(m_u+m_d)/2$.

A commonly reported measure of the strangeness sigma term is through
the $y$-parameter,
\begin{equation}
y\equiv\frac{2\langle N|\overline{s}s|N\rangle}{\langle N|\overline{u}u+\overline{d}d|N\rangle}
=\frac{2m_l}{m_s}\frac{\sigma_s}{\sigma_l}\,.
\end{equation}

\section{Phenomenological Determination}
\label{sec:pheno}
By the very nature of the weak coupling of the Higgs to the low-energy
sector of the Standard Model, the sigma terms are essentially
impossible to measure directly. Fortunately, $\sigma_l$ can be inferred
through a chiral low-energy relation, where the amount of explicit
chiral symmetry breaking can be related to pion--nucleon
scattering. In particular, the light-quark sigma term can be extracted
through measurement of the Born-subtracted, isoscalar amplitude
$\Sigma_{\pi N}(t)$.

Both $\Sigma_{\pi N}$ and $\sigma$ vanish with the quark mass, but
importantly they become equal as the chiral limit is
approached. That is, the leading dependence of the quark mass is the
same, with the leading difference being ${\mathcal O}(m_l^{3/2})$.  The limited
knowledge of this difference term can further be reduced by moving to
the unphysical kinematic point $t=2m_\pi^2$ (the Cheng-Dashen point
\cite{Cheng:1970mx}), where the remainder is ${\mathcal O}(m_l^2)$ \cite{Brown:1971pn},
being defined by
\begin{equation}
\Delta_R\equiv \Sigma_{\pi N}(t=2m_\pi^2)-\sigma_l(t=2m_\pi^2)={\cal O}(m_l^2)\,.
\end{equation}
Here, the scalar matrix element has been extended to non-zero momentum
transfer, with the usual sigma term corresponding to the $t\to 0$
limit, $\sigma_l=\sigma_l(t=0)$. An early calculation of the
remainder term has determined an estimate $\Delta_R\simeq 0.35\mev$
\cite{Gasser:1990ce}, later followed by an updated value
$\Delta_R\simeq 2\mev$ \cite{Bernard:1996nu}.

Extraction of the sigma term at $t=0$ then requires the determination
of the form factor correction, defined by
\begin{equation}
\Delta_\sigma\equiv \sigma_l(t=2m_\pi^2)-\sigma_l(t=0)\,.
\end{equation}
Through dispersion relations, this form factor correction has been
estimated to be $\Delta_\sigma\simeq 15\mev$ \cite{Gasser:1990ap}.

With the required theoretical corrections under reasonable control,
the pion--nucleon sigma term can then be extracted from the
experimentally determined $\Sigma_{\pi N}$, as extrapolated to the
unphysical Cheng-Dashen point. In summary, $\sigma_l$ is
determined by
\begin{equation}
\sigma_l= \Sigma_{\pi N}(t=2m_\pi^2)-\Delta_R-\Delta_\sigma.
\end{equation}
Following this outlined technique, analysis of $\pi N$ scattering data
gives the Gasser--Leutwyler--Sainio (GLS) value $\sigma_l=45\pm 8\mev$
\cite{Gasser:1990ce}, or a somewhat larger value from the George
Washington University/TRIUMF (GWU) group analysis $\sigma_l=64\pm
7\mev$ \cite{Pavan:2001wz}.

In addition to these benchmark calculations of GLS and GWU, an
analysis based on a covariant baryon chiral perturbation theory has
recently been reported by Alarc\'on, Martin-Camalich \& Oller (AMO)
\cite{Alarcon:2011zs}. Here the low-energy $\pi N$ scattering phase
shifts are fit directly within the effective field theory
framework. The analysis determines the relevant low-energy constants
necessary for the extraction of the sigma term by the
Hellmann--Feynman theorem (discussed below). Together with estimates
for the higher-order terms, the sigma term is reported to be
$\sigma_l=59\pm 7\mev$, lying between the two values reported above.

\begin{figure}
\begin{center}
\includegraphics[width=8cm]{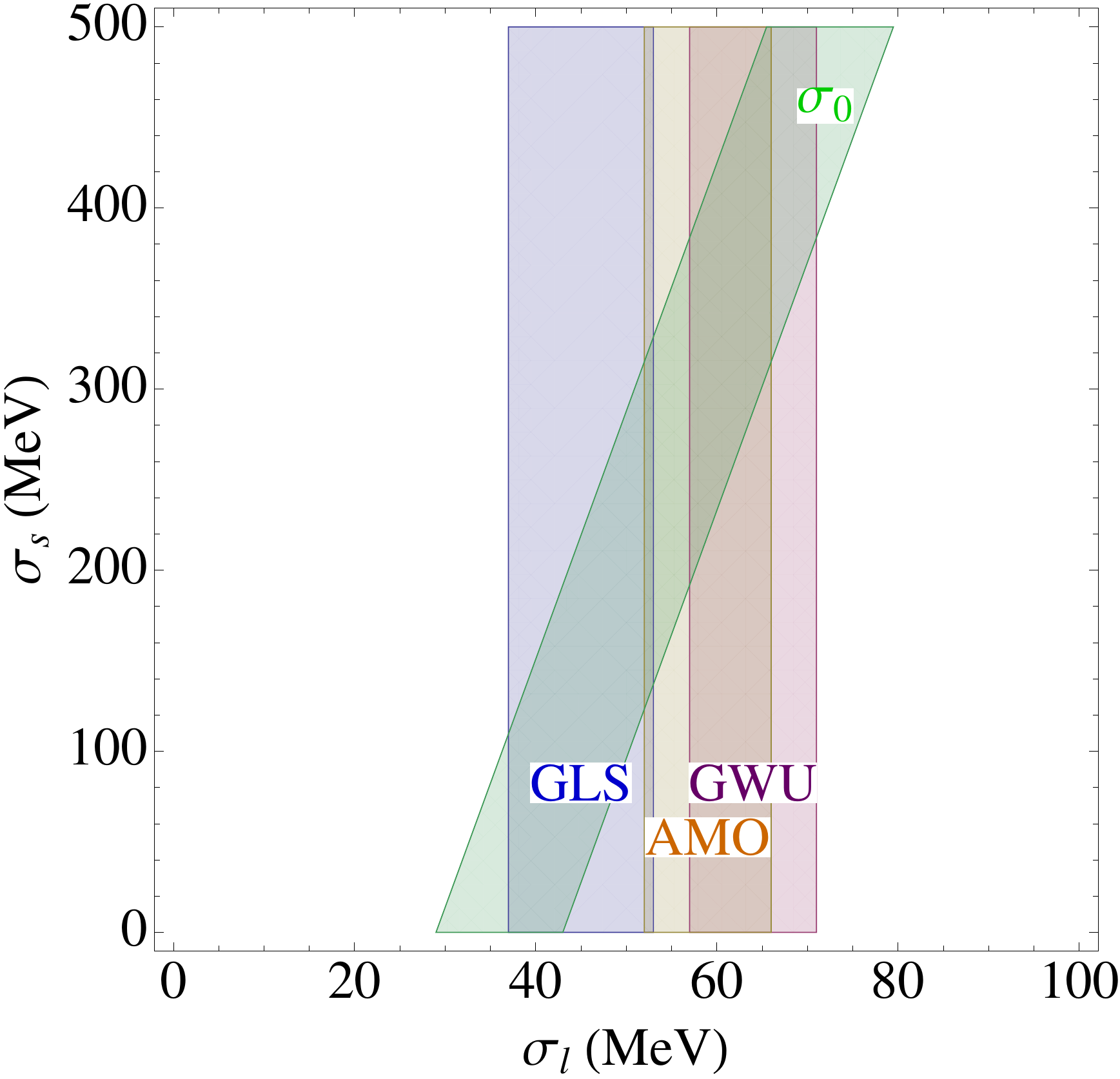}
\end{center}
\caption{The determination of $\sigma_s$, from the best-estimate of 
$\sigma_0=36\pm 7\mev$ \cite{Borasoy:1996bx}, has a high degree of 
sensitivity to the determination of pion--nucleon sigma term. The 
three values for $\sigma_l$ are as discussed in the text.
  \label{fig:sigma0}}
\end{figure}

Estimating the strange-quark sigma term is significantly more
challenging. In this case, the strange quark is too heavy to reliably
truncate the low-energy relation at low order. As an alternative, the
conventional approach has been to estimate $\sigma_s$ by studying the
SU(3) breaking among the baryon octet
\cite{Gasser:1980sb,Borasoy:1996bx}. Here, the baryon
mass splittings can be used to constrain the non-singlet combination
\begin{equation}
\sigma_0\equiv m_l\langle N|\overline{u}u+\overline{d}d-2\overline{s}s|N\rangle\,.
\end{equation}
To leading-order in the quark masses, $\sigma_0$ can be estimated from
the physically observed spectrum by
\begin{equation}
\sigma_0 \simeq \frac{m_l}{m_s-m_l}\left(M_\Sigma+M_\Xi-2M_N\right)\simeq 24\mev\,.
\end{equation}
By incorporating the higher-order terms in the quark mass expansion,
Borasoy and Mei\ss ner have determined an improved estimate
$\sigma_0=36\pm7\mev$ \cite{Borasoy:1996bx}.  With an estimate for
$\sigma_0$, the strangeness sigma term is then given by
\begin{equation}
\sigma_s=\frac{m_s}{2m_l}(\sigma_l-\sigma_0)\,.
\label{eq:sigs}
\end{equation}
Being multiplied by the large quark mass ratio, this method leads to a
value for $\sigma_s$ that is acutely sensitive to the difference between
$\sigma_l$ and $\sigma_0$. Further, given the limited precision
available to $\sigma_0$, even a perfect determination of $\sigma_l$
leaves a residual uncertainty in $\sigma_s$ of order $90\mev$. For the
three determinations of $\sigma_l$ discussed above,
Figure~\ref{fig:sigma0} displays the broad range of possible
$\sigma_s$ values.

Given the difficulty in extracting a precise determination of
$\sigma_s$ from phenomenology, there is significant scope for lattice
QCD to provide a meaningful constraint on this nucleon matrix
element. In addition, there is also the potential for lattice
simulations to shed light on the phenomenological determination of
$\sigma_l$.

\section{Lattice QCD}
Within the framework of lattice QCD, there are two main methods used
in the extraction of the sigma terms. These divide into the explicit
evaluation of the scalar $\qbar q$ matrix element by 3-point function
methods or by invoking the Hellmann--Feynman relation through the study of
the quark-mass dependence of the nucleon mass.

\begin{figure}
\begin{center}
\includegraphics[width=8cm]{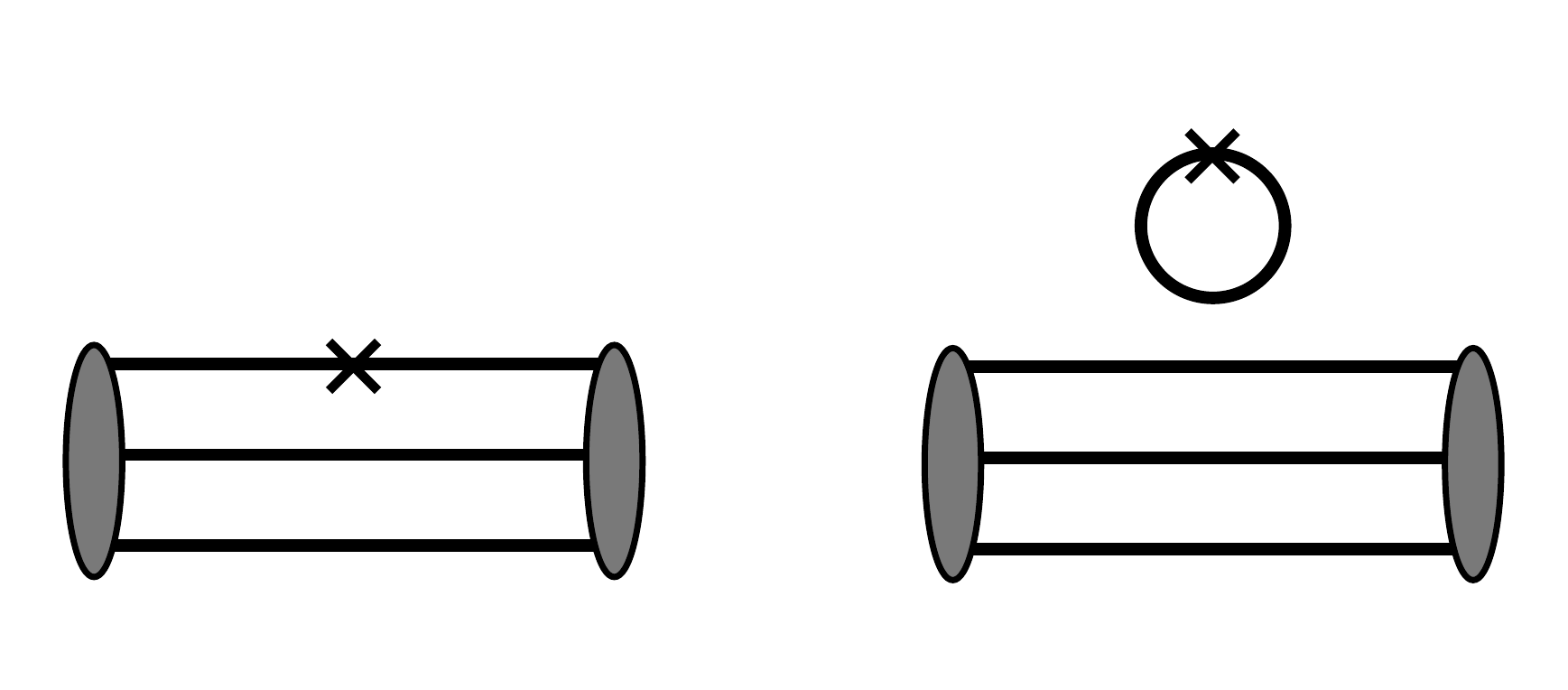}
\end{center}
\caption{Schematic diagram displaying the two topologically distinct contractions
  in the evaluation of the scalar 3-point function. The standard
  jargon is that the left is a {\it connected} insertion and the right
  a {\it disconnected} insertion.
\label{fig:3pt}}
\end{figure}

In the first technique, appropriate ratios of 3-point and 2-point
correlation functions are formed to isolate the matrix elements of
interest. The evaluation of the correlation functions involves two
distinct forms of Wick contraction, as depicted in Figure
\ref{fig:3pt}. In particular, there are contributions involving quark
line connected or disconnected operator insertions. While standard
methods typically yield strong signals for the connected correlation
functions, it is well-known that the disconnected insertions are
notoriously challenging, see eg.~Refs.~\cite{Bali:2009hu,Dinter:2012tt,Babich:2010at}. This of particular
significance for the strange-quark matrix elements in the nucleon,
which are purely disconnected.

A common alternative to the three-point method, is to determine the
$\qbar q$ matrix elements by differentiation with respect to the quark
masses, where the Hellmann--Feynman relation gives
\cite{Hellmann:1937,Feynman:1939zza,Gasser:1979hf}
\begin{equation}
\sigma_q=m_q\frac{\del M_N}{\del m_q}\,.
\end{equation}
Here the Gell-Mann--Oakes--Renner (GOR) relation
\cite{GellMann:1968rz} is commonly imposed such that $M_N$ is
expressed as a function of the squares of meson masses. One way to do
this is to write $M_N=M_N(m_\pi^2,\tilde{m}_K^2)$, where
$\tilde{m}_K^2=m_K^2-m_\pi^2/2$ is the projection of the square of the
kaon mass onto the SU(2) chiral limit ($m_l\to 0$). With such a
formulation, the sigma terms (for 2+1-flavour simulations) are easily
written as
\begin{equation}
\sigma_l = m_\pi^2 \frac{\partial M_N}{\partial m_\pi^2}\,,\qquad
\sigma_s = \tilde{m}_K^2 \frac{\partial M_N}{\partial \tilde{m}_K^2}\,.
\end{equation}
The main challenge of this approach is the difficulty in reliably
parameterising the quark-mass dependence over a range of light and
strange quark masses. In particular, it is only in recent years that
there have been large scale numerical simulations of baryons with
2+1-flavours of dynamical quarks,
eg.~\cite{Bernard:2001av,WalkerLoud:2008bp,Aoki:2008sm,Lin:2008pr,Durr:2008zz,Baron:2010bv,Bietenholz:2011qq}. In
constraining the two-dimensional parameter space, it is also the case
that typical lattice trajectories in the $m_l$--$m_s$ plane approach the
physical point for approximately constant $m_s$; though the
QCDSF-UKQCD are currently pursuing an approach which keeps the singlet
combination $(2m_l+m_s)$ a constant \cite{Bietenholz:2011qq}.

\begin{figure}
\begin{center}
\includegraphics[width=10cm]{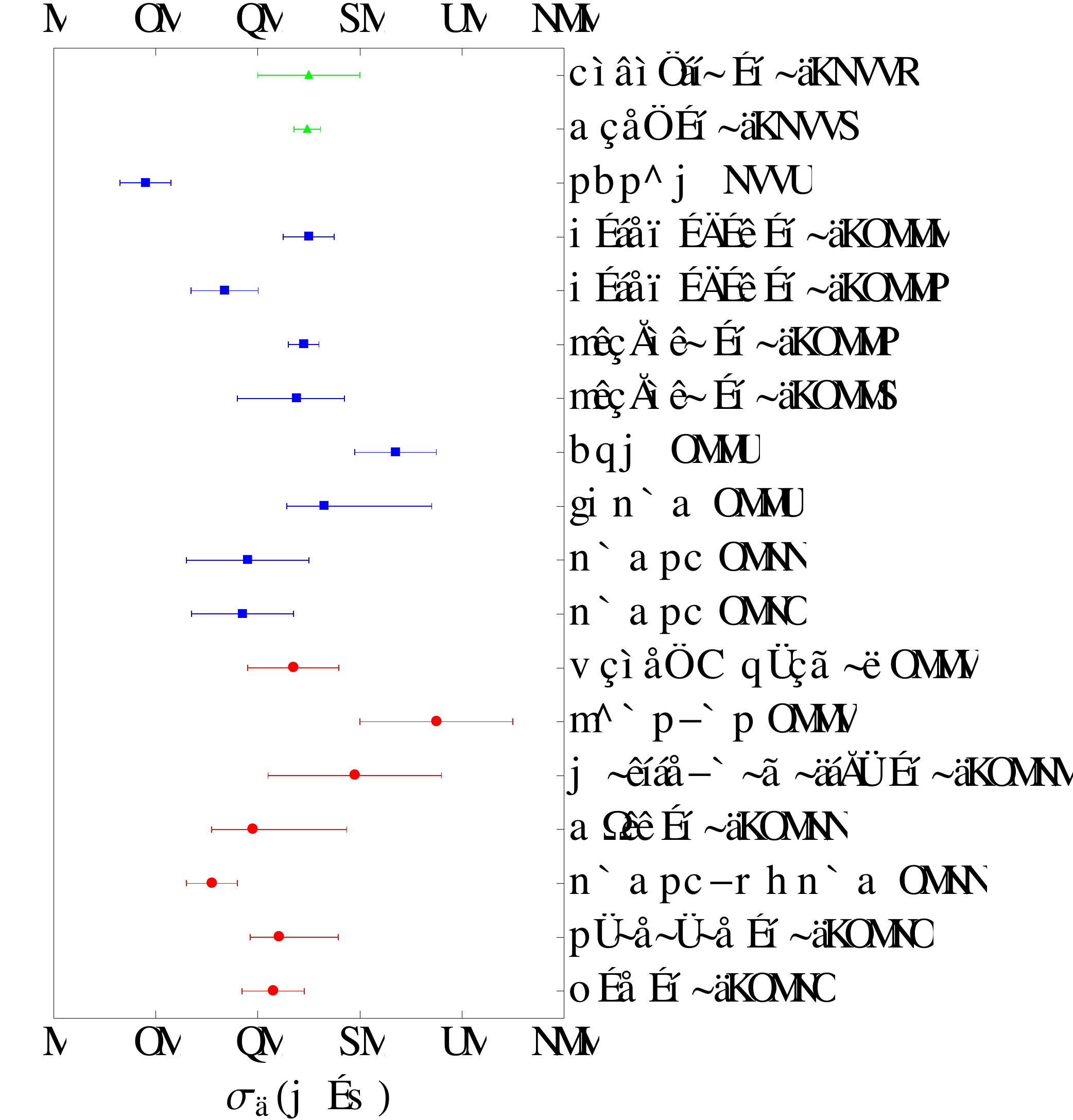}
\end{center}
\caption{Light-quark sigma term results based on lattice QCD. The colours denoted the number of dynamical flavours of quarks: green is $N_f=0$, blue $N_f=2$ and red $N_f\ge 2+1$. References: Fukugita
  \etal \cite{Fukugita:1994ba}, Dong \etal \cite{Dong:1995ec}, SESAM
  \cite{Gusken:1998wy}, Leinweber \etal (2000) \cite{Leinweber:2000sa},
  Leinweber \etal (2003) \cite{Leinweber:2003dg}, Procura \etal (2003)
  \cite{Procura:2003ig}, Procura \etal (2006) \cite{Procura:2006bj}, ETM
  \cite{Alexandrou:2008tn}, JLQCD \cite{Ohki:2008ff}, QCDSF (2011)
  \cite{Bali:2011ks}, QCDSF (2012) \cite{Bali:2012qs}, Young \& Thomas
  \cite{Young:2009zb}, PACS-CS \cite{Ishikawa:2009vc}, Martin-Camalich
  \etal \cite{MartinCamalich:2010fp}, D\"urr \etal \cite{Durr:2011mp},
  QCDSF-UKQCD \cite{Horsley:2011wr}, Shanahan \etal
  \cite{Shanahan:2012wh}, Ren \etal \cite{Ren:2012aj}.
  \label{fig:sigl} }
\end{figure}

\begin{figure}
\begin{center}
\includegraphics[width=10cm]{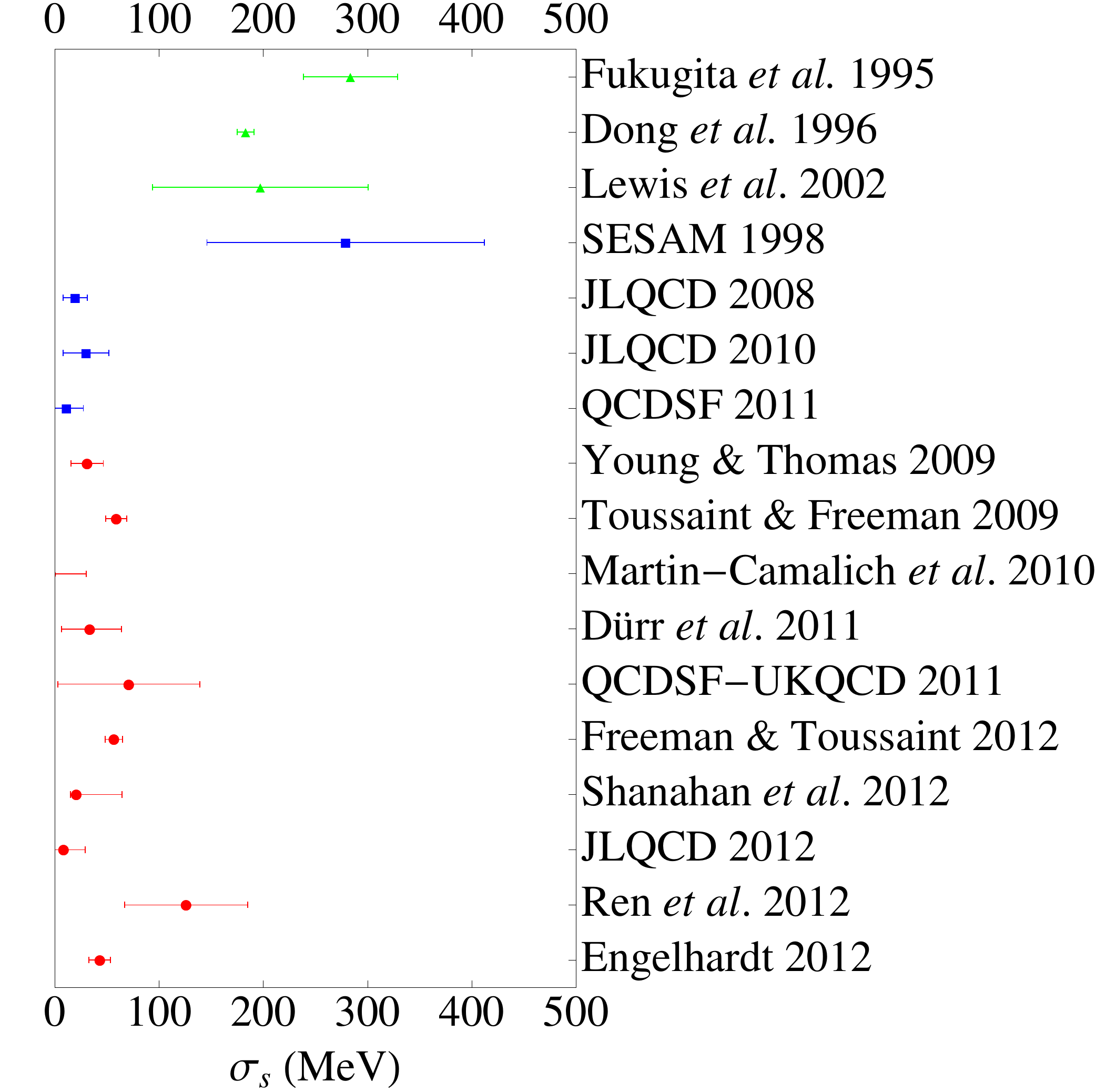}
\end{center}
\caption{Strange-quark sigma term results based on lattice QCD. Colours as in Figure~3.
  Fukugita \etal \cite{Fukugita:1994ba}, Dong \etal
  \cite{Dong:1995ec}, Lewis \etal \cite{Lewis:2002ix}, SESAM
  \cite{Gusken:1998wy}, JLQCD (2008) \cite{Ohki:2008ff}, JLQCD 2010
  \cite{Takeda:2010cw}, QCDSF \cite{Bali:2011ks}, Young \& Thomas
  \cite{Young:2009zb}, Toussaint \& Freeman \cite{Toussaint:2009pz},
  Martin-Camalich \etal \cite{MartinCamalich:2010fp}, D\"urr \etal
  \cite{Durr:2011mp}, QCDSF-UKQCD \cite{Horsley:2011wr}, Freeman \&
  Toussaint \cite{Freeman:2012ry}, Shanahan \etal
  \cite{Shanahan:2012wh}, JLQCD (2012) \cite{Ohki:2012rzb}, Ren \etal
  \cite{Ren:2012aj}, Engelhardt \cite{Engelhardt:2012gd}.
  \label{fig:sigs}}
\end{figure}

\begin{figure}
\begin{center}
\includegraphics[width=10cm]{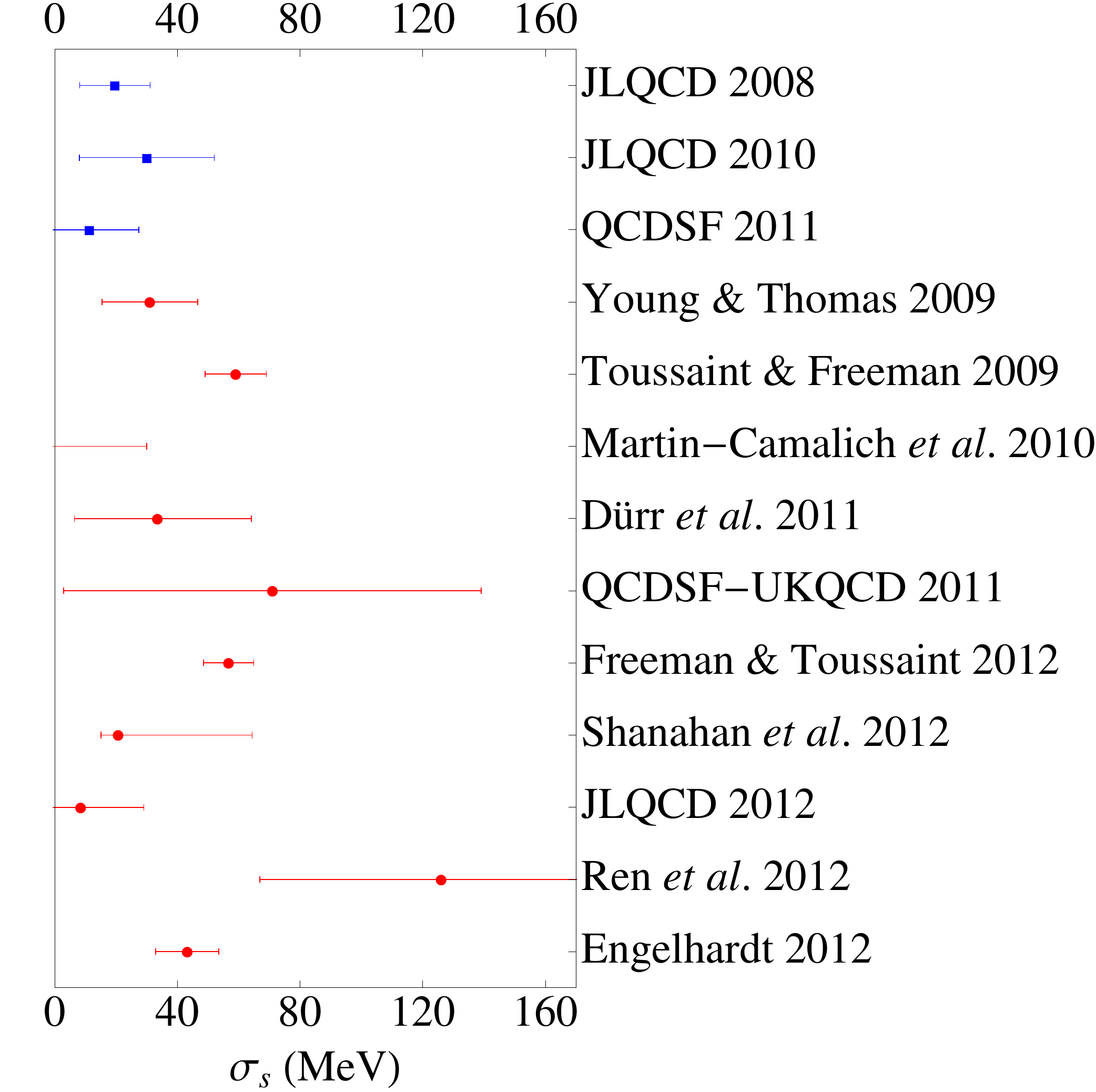}
\end{center}
\caption{Zoomed in graphic of Fig.~2 showing more recent results on $\sigma_s$.
\label{fig:zoom}}
\end{figure}

A summary of the progression of lattice results is displayed in
Figures \ref{fig:sigl} and \ref{fig:sigs}.
With differing degrees of analysis into the various uncertainties of
the calculations, it would be difficult to formulate any rigourous
aggregates. Nevertheless, two clear features are emergent. For
$\sigma_l$, the values revealed in lattice simulations are compatible
with the range of phenomenologically determined values. Secondly,
the modern values of $\sigma_s$ are all at the lower end of the
possible values suggested by Figure~\ref{fig:sigma0}. To highlight the
current status, Figure~\ref{fig:zoom} shows a close-up of the recent
determinations of $\sigma_s$.

\section{Dark Matter}
The smaller values of $\sigma_s$ revealed in the recent lattice
studies are particularly significant in the context of the direct
search for dark matter. The most precise limits on WIMP--nucleon cross
sections are being constrained by the XENON100 Collaboration, with the 
latest update placing an upper bound on the cross section of less than
$10^{-44}\cm^2$ over a wide range of WIMP masses
\cite{Aprile:2012nq}. Figure~3 of \cite{Aprile:2012nq} suggests these
limits are continuing to reduce the parameter space of potential
supersymmetric candidates for dark matter. 

The XENON100 Collaboration results are plotted against predicted cross sections
for some favoured supersymmetric models
\cite{Strege:2011pk,Fowlie:2012im,Buchmueller:2011ab}. The predicted
cross section rates are based on a determination of the strange quark
sigma term, $\sigma_s$,\footnote{Or in an alternative common notation,
  $f_{Ts}=\sigma_s/M_p$, for the proton mass $M_p$.} as outlined in
Section~\ref{sec:pheno}. Hence $\sigma_s$ in these studies exhibits
the extreme sensitivity to $\sigma_l$ displayed in
Figure~\ref{fig:sigma0}.

\begin{figure}
\begin{center}
\includegraphics[width=10cm]{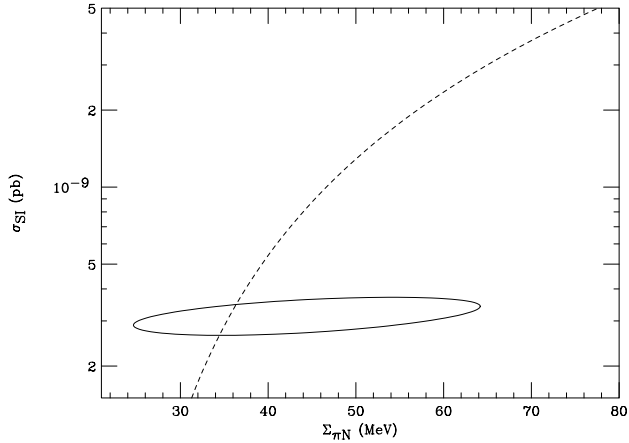}
\end{center}
\caption{
\label{fig:cmssm}
For a pre-LHC CMSSM dark matter model, the predicted spin-independent
cross section ($\sigma_{\rm SI}$) shows a strong dependence on
$\Sigma_{\pi N}$. This variation is a consequence of the large
variation of $\sigma_s$, as constrained by the phenomenological
$\sigma_0$, as seen in Figure~1.
The ellipse
show the range of $\sigma_{\rm SI}$ using lattice inputs for
$\Sigma_{\pi N}$ and $\sigma_s$ \cite{Giedt:2009mr}.}
\end{figure}

As the WIMP--nucleon interactions are largely Higgs-coupling driven,
the difference between a small and large $\sigma_s$ can have a
dramatic influence on the predicted cross sections. This is
highlighted in Figure~\ref{fig:cmssm}, which shows how the predicted
cross section for a particular constrained minimal supersymmetric
standard model (CMSSM) model\footnote{The figure displays the
  predicted cross-section for ``model C'', as one of a class of
  benchmark models proposed pre-LHC
  \cite{Battaglia:2001zp,Battaglia:2003ab}.} depends strongly on
$\Sigma_{\pi N}$ (with $\sigma_s$ constrained by the phenomenological
$\sigma_0$) \cite{Giedt:2009mr}.
In contrast, the displayed ellipse shows the range of predicted cross
sections within the 95\% confidence level interval of the lattice QCD
determinations of $\sigma_l$ and $\sigma_s$ from
Refs.~\cite{Young:2009zb,Toussaint:2009pz}. It should be stressed that
the reduced variation in the cross section is a consequence of the
increased precision in $\sigma_s$ from lattice QCD input --- which is
not reliant on the propogation of the phenomenological uncertainty in
$\sigma_0$.

Generic dark matter cross section packages, such as micrOMEGAs
\cite{Belanger:2008sj}, have been designed to take as inputs
$\sigma_l$ and $\sigma_0$.
With the improvement in lattice QCD results discussed above, it 
would be advantageous to see these packages reformulated to take
$\sigma_l$ and $\sigma_s$ as inputs\footnote{Of course these are precisely the same thing with an
appropriately included correlation coefficient.}. In the meantime,
with cross section predictions based on $\sigma_0$ as
an input, the reduction in uncertainty in $\sigma_s$ may be
equivalently stated as a reduction in $\sigma_l-\sigma_0$,
cf.~Eq.~(\ref{eq:sigs}). A crude, yet conservative view of
Figure~\ref{fig:zoom}, may suggest a value\footnote{The final
  numerical value uses the central estimate of the quark mass
  ratio from the FLAG \cite{Colangelo:2010et}, which is primarily
  determined by the precision studies of
  Refs.~\cite{Bazavov:2009bb,Durr:2010vn}.}
\begin{equation}
\sigma_l-\sigma_0=\frac{2m_l}{m_s}(40\pm 30\mev)\simeq 2.9\pm 2.2\mev\,.
\end{equation}
Already at this scale of precision, there should be a substantial
reduction in the uncertainties in $\sigma_{\rm SI}$ associated with the
hadronic matrix elements. Importantly, with the reduction in the
hadronic uncertainty, any discovery of dark matter will have
significantly more discrimination power among candidate models.

\section{Summary}
To summarise, an accurate determination of the strange quark sigma term is
of principle importance in the reduction of hadronic uncertainties in
the predicted dark matter cross sections for a wide range of models. In
the determination of the relevant nucleon scalar matrix elements,
lattice QCD simulations have made significant progress in recent years
--- particularly with the emergence of dynamical simulations with 2+1
flavours of dynamical quarks. As a confirmation of lattice
methods, it is reassuring to observe that the pion--nucleon sigma term
appears to be compatible with the reliably determined phenomenological
extraction. Further, recent determinations of the strange quark
sigma term are significantly smaller than had previously been suggested
--- and lattice calculations are now at a far greater precision.
With the prospect of a discovery of dark matter in the near future, it
will be essential for lattice QCD simulations to further reduce these
hadronic uncertainties.

\section*{Acknowledgements}
I thank my collaborators for their contributions to various aspects of
the work presented here, J.~Giedt, P.E.~Shanahan, A.W.~Thomas and
S.J.~Underwood.
This work was supported by the Australian Research Council through the
ARC Centre of Excellence for Particle Physics at the Terascale and
grants DP110101265 and FT120100821.

\end{document}